\documentclass[prl,twocolumn,amsmath,amssymb,graphicx]{revtex4-2}

\usepackage{bm,url,natbib,textcase,color}

\begin{document}

\title{Vertex-Constraints in 3D Higher Spin Theories}

\author{Stefan Fredenhagen}
\email{stefan.fredenhagen@univie.ac.at}
\affiliation{University of Vienna, Faculty of Physics, Boltzmanngasse 5, 1090 Vienna, Austria}
\affiliation{Erwin Schr\"odinger International Institute for Mathematics and Physics, Boltzmanngasse 9, 1090 Vienna, Austria}
\author{Olaf Kr\"uger}
\email{olaf.krueger@univie.ac.at}
\affiliation{University of Vienna, Faculty of Physics, Boltzmanngasse 5, 1090 Vienna, Austria}
\author{Karapet Mkrtchyan} 
\email{karapet.mkrtchyan@aei.mpg.de}
\affiliation{Max Planck Institute for Gravitational Physics (Albert Einstein Institute),
Am M\"uhlenberg 1, 14476 Potsdam, Germany}

\date{\today{}}

\begin{abstract}
  We analyse the constraints imposed by gauge invariance on higher-order interactions between massless bosonic
  fields in three-dimensional higher-spin gravities. We show that vertices of quartic and higher order that are
  independent of the cubic ones can only involve scalars and Maxwell fields. As a consequence, the full
  non-linear interactions of massless higher-spin fields are completely fixed by the cubic vertex.
\end{abstract}

\maketitle

\section{Introduction}

In this Letter, we start an investigation aimed at a Lagrangian formulation of three-dimensional (3D)
higher-spin (HS) gravities \cite{Prokushkin:1998bq} beyond cubic order.

HS gravity theories are generalisations of gravity, where higher-spin gauge fields are introduced. In 3D, a free
spin-$s$ gauge field is a symmetric tensor field $\phi_{\mu_{1}\dots \mu_{s}}$ with gauge transformation
\begin{equation}\label{freegauge}
  \delta^{(0)} \phi_{\mu_{1}\dots \mu_{s}}= s\, \partial_{(\mu_{1}}\epsilon_{\mu_{2}\dots \mu_{s})}\,,
\end{equation}
similar to Maxwell or Chern-Simons vector gauge fields ($s=1$) and linearised gravity ($s=2$). It is described
by the quadratic Fronsdal Lagrangian $\mathcal{L}_{2}$ \cite{Fronsdal:1978rb}. We collectively denote massless
fields with spin $s > 1$ and Chern-Simons vector fields as ``massless HS fields''. In 3D, these do not possess
propagating degrees of freedom (d.o.f.), however, they can have interesting boundary dynamics at the conformal
boundary of asymptotically Anti-de Sitter (AdS) space-times. Up to now, no non-linear Lagrangian of interacting
Fronsdal fields is known, but there is a systematic perturbative approach to construct such Langrangians. This
is known as the Noether-Fronsdal program, which we follow in this work and review below.

There are different motivations to study HS gravities. Most prominently, they constitute generalisations of
gravity for which holographic dualities can be investigated: a $(d+1)$-dimensional HS gravity theory on
asymptotically AdS space-times is related to a $d$-dimensional conformal field theory (CFT). This HS AdS/CFT
correspondence \cite{Klebanov:2002ja,Giombi:2012ms} is a priori independent of the string-theoretic AdS/CFT
correspondence, and possesses distinct features as it does not require supersymmetry and is accessible to
perturbative checks. It becomes particularly interesting for 3D HS gravities \cite{Gaberdiel:2010pz}, because
for 2D CFTs many exact results are available. These also allow to study the relation between the tensionless
limit of string theory and HS theories via their CFT dual \cite{Gaberdiel:2014cha,Gaberdiel:2015wpo}.

To perform computations on the HS side, finding a Lagrangian formulation is crucial. For the non-propagating
sector (i.e.\ without scalars or Maxwell fields), a non-linear action is available in Chern-Simons form
\cite{Blencowe:1988gj,Campoleoni:2010zq,Afshar:2013vka} (which is a generalisation of the Chern-Simons
formulation of 3D gravity \cite{Achucarro:1987vz}). There, one uses the frame-like formulation of HS fields in
terms of generalised vielbein fields and spin connections. In this formulation, coupling to matter is not
straightforward. It can be achieved by following the Vasiliev approach \cite{Prokushkin:1998vn} which uses
infinitely many auxiliary fields and for which no standard action is known.

The metric-like formulation of HS gravity
\cite{Campoleoni:2012hp,Fredenhagen:2014oua,Mkrtchyan:2017ixk,Kessel:2018ugi} (based on Fronsdal fields) is more
suitable for matter coupling. For example, the cubic interactions of massless HS fields are well studied both in
flat
\cite{Bengtsson:1986kh,Metsaev:2005ar,Manvelyan:2010wp,Manvelyan:2010jr,Sagnotti:2010at,Fotopoulos:2010ay,Manvelyan:2010je,Mkrtchyan:2011uh,Metsaev:2012uy,Henneaux:2013gba,Conde:2016izb}
and $(A)dS$ spaces
\cite{Fradkin:1986qy,Vasilev:2011xf,Joung:2011ww,Manvelyan:2012ww,Boulanger:2012dx,Francia:2016weg} of
dimensions $D\geq 4$. However, the main challenge in formulating the action in arbitrary dimensions arises at
quartic order (see, e.g.,
\cite{Berends:1984rq,Bekaert:2015tva,Bengtsson:2016hss,Roiban:2017iqg,Sleight:2017pcz,Ponomarev:2017qab}) and
this is also expected in the 3D case with matter.

In the Noether procedure one starts with the free quadratic Lagrangian $\mathcal{L}_{2}$ and builds vertices
order by order, including matter couplings. For a given HS theory, we expand the Lagrangian in powers of small
parameters $g_n$,
\begin{equation*}
  \mathcal{L}=\mathcal{L}_{2} + \sum_{n\geq 3} g_{n} \mathcal{L}_n + O(g_{n}^2)\,,
\end{equation*}
where we suppress a sum over the different kinds of $n$-point vertices $\mathcal{L}_n$. Altogether,
$\mathcal{L}$ must be gauge invariant, $\delta \mathcal{L}=0$, up to boundary terms, where $\delta$ is obtained
by deforming the transformation of the free fields (see Eq.~(\ref{freegauge})),
\begin{equation*}
  \delta=\delta^{(0)}+\delta^{(1)}+\dots\,,
\end{equation*}
expanded in powers of the fields.

Cubic gauge invariant vertices in 3D have been classified \cite{Mkrtchyan:2017ixk,Kessel:2018ugi}. In this work,
we study higher-order vertices of massless fields that are independent of the ones of lower order. Because of
gauge invariance, they satisfy the following Noether equations:
\begin{equation}\label{NE}
  \delta^{(n-2)} \mathcal{L}_{2}+\delta^{(0)}\mathcal{L}_{n} = 0\ \ \text{up to total derivatives}\,.
\end{equation}
We show that after suitable field redefinitions \textit{in 3D all such vertices of order $n \geq 4$ contain no
  massless higher-spin fields}.

\section{Preliminaries}

The Lagrangian $\mathcal{L}$ is written in terms of massless Fronsdal fields, subject to non-linear gauge
transformations. For the classification purposes, we are only interested in the part of the vertices that do not
contain divergences and traces of the fields, even though the traceless and transverse (TT) condition on
Fronsdal fields is not achieved by off-shell gauge fixing. Hence, from now on we assume that the fields are
parametrised by symmetric, traceless and divergence-free tensors $\phi_{\mu_1\dots\mu_s}(x)$ with
$\mu_i \in (0,1,2)$; $s$ denotes the spin of the field and the corresponding free equation of motion (e.o.m.) is
the Klein-Gordon equation with zero mass (see, e.g., \cite{Kessel:2018ugi}).

For convenience, one contracts the tensor indices each with an auxiliary vector variable $a^\mu$. This defines
\begin{equation}
  \label{eq:phi(a)}
  \phi^{(s)}(x,a)=\frac{1}{s!}\phi_{\mu_1\dots\mu_s}(x) a^{\mu_1}\cdots a^{\mu_s}\,
\end{equation}
and the properties of $\phi_{\mu_1\dots\mu_s}(x)$ translate to the \textit{Fierz equations} for
$\phi^{(s)}(x,a)$:
\begin{equation*}
  A^2 \;\phi^{(s)} =  A \cdot P\; \phi^{(s)} =  P^2\; \phi^{(s)}\big|_{\text{free e.o.m.}} = 0\,,
\end{equation*}
where $P^\mu = \partial_{x^\mu}$ and $A^\mu = \partial_{a^\mu}$.

We analyse the general form of the deformations $\mathcal{L}_n$ for $n\geq 4$, which can be written as
\begin{equation}
  \label{eq:L^n}
  \mathcal{L}_n = \mathcal{V} \left( \prod_{i = 1}^n \phi_i(x_i,a_i) \right) \Bigg|_{\substack{x_i = x\\ a_i = 0}}\,,
\end{equation}
where we abbreviated $\phi_i = \phi^{(s_i)}$. The \textit{vertex generating operator} $\mathcal{V}$ performs the
index contractions via the operators $P^\mu_i = \partial_{x^\mu_i}$ and $A^\mu_i = \partial_{a^\mu_i}$. Let us
first concentrate on \textit{parity even vertices} $\mathcal{L}_n$, hence $\mathcal{V}$ is a polynomial in the
commuting variables
\begin{equation*}
  z_{ij} = A_i \cdot A_j , \quad y_{ij} = A_i \cdot P_j  \quad s_{ij} = P_i \cdot P_j\,.
\end{equation*}
These contract two indices each: One from $\phi_i$ with one from $\phi_j$ ($z_{ij}$); one from $\phi_i$ with one
from a derivative acting on $\phi_j$ ($y_{ij}$); and two from derivatives acting on $\phi_i$ and $\phi_j$
($s_{ij}$). The $s_{ij}$ are the familiar \textit{Mandelstam variables}. In the end, we set $a_i = 0$ to ensure
Lorentz invariance. Whenever appropriate, we use an $n$-periodic index notation, e.g. $s_{i\, n+j} = s_{ij}$.

\section{Equivalence Relations}

We say that two vertex generating operators $\mathcal{V}$ and $\mathcal{V}'$ are equivalent,
$\mathcal{V} \approx \mathcal{V}'$, if and only if the two resulting Lagrangians $\mathcal{L}_n$ and
$\mathcal{L}_n'$, constructed via Eq.~(\ref{eq:L^n}), describe the same theory. Hence, we seek the most general
form of $\mathcal{V}$, \textit{up to equivalence}. 

For example, field redefinitions $\phi_i \mapsto \phi_i + \delta \phi_i$, such that $\delta \phi_i$ is
non-linear in the fields, do not alter the theory, but may affect the form of the vertices $\mathcal{L}_n$. This
freedom of field redefinitions can be used to choose $\mathcal{V}$ to be independent of $s_{ii}$. This
generalises the so-called Metsaev basis for cubic vertices
\cite{Kessel:2018ugi,Manvelyan:2010wp,Manvelyan:2010jr,Conde:2016izb} to higher $n$.

Since we are interested in the TT part of the vertex, $\mathcal{V}$ does not depend on $z_{ii}$ and
$y_{ii}$. 
So far, we summarize that $\mathcal{V}$ is an element in the polynomial ring
$\mathcal{R} = \mathbb{R}[z_{ij}|_{i < j}, y_{ij}|_{i \neq j}, s_{ij}|_{i < j}]$.

Furthermore, acting with $D^\mu = \sum_{j = 1}^n P_j^\mu$ on the expression in brackets in Eq.~(\ref{eq:L^n})
results in a total derivative term in the Lagrangian which does not affect the theory. We may hence remove any
dependence of $\mathcal{V}$ on $A_i \cdot D$ and $P_i \cdot D$. In other words, we impose the equivalence
relations
\begin{equation*}
  \sum_{j = 1}^n y_{ij} \approx 0\,,\quad  \sum_{j = 1}^n s_{ij} \approx 0 \,,
\end{equation*}
which generate an ideal $\mathcal{I}_D \subset \mathcal{R}$.

A final class of equivalence relations is given by Schouten identities, which stem from over-antisymmetrisation
of space-time indices within the Lagrangian. They translate to an ideal $\mathcal{I}_S$ of equivalence relations
in $\mathcal{R}$ as follows: Consider the vector of derivative operators
$b = ( P_1 , \ldots , P_n, A_1, \ldots A_n )$ and the symmetric $2n \times 2n $ matrix
\begin{equation*}
  \mathcal{B} = 
  \big(
  b_K \cdot b_L 
  \big)\big|_{K,L \in (1,\ldots, 2n)} = 
  \begin{pmatrix}
    \mathcal{S} & \mathcal{Y}^T \\ \mathcal{Y} & \mathcal{Z}
  \end{pmatrix}\,.
\end{equation*}
Here, $\mathcal{S} = (s_{ij})$, $\mathcal{Y} = (y_{ij})$, $\mathcal{Z} = (z_{ij})$ are $n\times n$ matrices with
elements in $\mathcal{R}$ (hence, their diagonal elements vanish). Now, remove $2n - 4$ rows and columns from
$\mathcal{B}$ and call the resulting $4\times 4$ matrix $M$. Acting with $\det M$ on the term in brackets in
Eq.~(\ref{eq:L^n}) yields an expression with four antisymmetrized space-time indices, which vanishes in three
dimensions. All such $4\times4$ minors of $\mathcal{B}$ form a generating set for the ideal $\mathcal{I}_S$.

All in all, $\mathcal{V}$ is a representative of an equivalence class in the quotient ring
$[\mathcal{V}] \in \mathcal{R}/(\mathcal{I}_D + \mathcal{I}_S)$ and we are free to choose a convenient one,
since all generating operators in one equivalence class describe the same vertex. However, it is hard to find
simple representatives, because the ideal $\mathcal{I}_S$ is too complicated. In the next section, we show that
it is easier to get a hold on representatives of $[\Delta \mathcal{V}]$, where we multiply $\mathcal{V}$ by an
appropriate product $\Delta$ of Mandelstam variables $s_{ij}$. The operator $\Delta \mathcal{V}$ correponds to
acting with contracted space-time derivatives on the vertex generated by $\mathcal{V}$. Then, by choosing a
simple representative for $[\Delta \mathcal{V}]$, we can impose strong constraints on the vertex generating
operator $\mathcal{V}$ itself. We show this shortly.

For the rest of this section, we show the following, essential observation: if $\Delta \mathcal{V}$ corresponds
to a trivial vertex, then the same is true for $\mathcal{V}$,
\begin{equation}
  \label{eq:main}
  \Delta\mathcal{V} \approx 0 \quad \Longrightarrow \quad \mathcal{V} \approx 0\,.
\end{equation}
This can be seen in Fourier space, where the operators $s_{ij}$ can be treated as numbers. If $\Delta$ is a
product of $s_{ij}$ ($i\not= j$), then it is generically non-zero on the subvariety in $k$-space defined by
$k_{i}^{2}=0$ and $\sum k_{i}=0$. The property $\Delta\mathcal{V} \approx 0$ translates in Fourier space to the
condition that $\Delta\mathcal{V}$ applied on any product of fields $\widehat{\phi}_{i}(k_{i},a_{i})$ (evaluated
at $a_{i}=0$) vanishes on this subvariety. As $\Delta$ is non-vanishing almost everywhere and $\mathcal{V}$ only
depends polynomially on $k_{i}^{\mu}$, one concludes that $\mathcal{V}$ applied on the fields
$\widehat{\phi}_{i}$ vanishes, hence $\mathcal{V}\approx 0$.

\section{Choice of Representatives}

In this section, we multiply a given vertex generating operator $\mathcal{V}$ with an appropriate product
$\Delta$ of Mandelstam variables $s_{ij}$ and choose a convenient representative for $[\Delta\mathcal{V}]$.
First, let $M$ be a $4\times4$ submatrix of $\mathcal{B}$ including the first three rows and columns, as well as
the $(n+i)$th row and $(n+j)$th column with $i \neq j$. Using the corresponding Schouten identity
$\det M \approx 0$, we can replace $z_{ij}$ in $\Delta \mathcal{V}$ by the $y_{kl}$ and $s_{kl}$ variables.
Doing this for all pairs $(i \neq j)$ allows us to choose a representative for $[\Delta \mathcal{V}]$ that does
not depend on $z_{ij}$ (we assume that $\Delta$ is chosen accordingly). Secondly, pick out a $4\times4$
submatrix $M$ of $\mathcal{B}$ including the columns $i$, $i+1$, $i+2$ (modulo $n$) and $(n+i)$, such that the
latter contains the elements $y_{ii} = 0$, $y_{i i+1}$, $y_{i i + 2}$ and any other, say $y_{i j}$. The Schouten
identities $\det M \approx 0$ can be used to replace all of the operators $y_{ij}$ in $\Delta \mathcal{V}$ by
$y_{i\,i+1}$, $y_{i\,i+2}$ and the Mandelstam variables. Finally, we perform a change of variables by replacing
each $y_{ii+2}$ in $\Delta \mathcal{V}$ by a linear combination of $y_{ii+1}$ and
$Y_i := s_{ii+2} y_{ii+1} - s_{ii+1} y_{ii+2}$. The reason for this replacement becomes more apparent in the
next section, but note for now that $Y_i^2 \approx 0$ due to Schouten identities. Indeed, the $4\times4$ minor
$\det M$ of $\mathcal{B}$, which consists of the rows and columns $i$, $i+1$, $i+2$ (modulo $n$) and $i+n$
satisfies $\det M = Y_i^2$.

We conclude that for a given vertex generating operator $\mathcal{V}$, there exists a product of Mandelstam
variables $\Delta$, such that
\begin{equation}
  \label{eq:P}
  \Delta \mathcal{V} \approx Q_\mathcal{V} (y_{ii+1} , Y_i , s_{ij})\,,
\end{equation}
and the polynomial $Q_\mathcal{V}$ is at most \textit{linear} in each $Y_i$ (a term $Y_iY_j$ with $i \neq j$ is
still possible, but $Y_i^2$ is not). We note that the polynomial might not be unique. It can be seen as a
representative of an equivalence class
\begin{equation*}
  [Q_\mathcal{V}] \in \frac{\mathbb{R}[y_{ii+1}, Y_i , s_{ij}]}{\mathcal{I}_R + \langle Y_i^2 \rangle}\,, 
\end{equation*}
where the ideal $\mathcal{I}_R \subset \mathbb{R}[y_{ii+1}, Y_i , s_{ij}]$ is generated by all remaining
equivalence relations (total derivatives and Schouten identities).

\section{Constraints from Gauge Invariance}

In this section, we show that gauge invariance implies that the polynomial $Q_\mathcal{V}$ in Eq.~(\ref{eq:P})
does not depend on $y_{ii+1}$. To this end, we consider the 0th order gauge transformations of the fields (see
Eq.~(\ref{freegauge})),
\begin{equation*}
  \delta^{(0)}\phi^{(s)}(x, a)= a \cdot P \,\epsilon^{(s-1)}(x, a)\,,
\end{equation*}
where the gauge parameter $\epsilon^{(s-1)}$, constructed as in Eq.~(\ref{eq:phi(a)}), also satisfies the Fierz
equations.

In the condition for gauge invariance, Eq.~(\ref{NE}), the first term vanishes when the free e.o.m.\ for the
fields are applied. Hence,
\begin{equation*}
  \delta^{(0)}_k \mathcal{L}_n = \mathcal{V} \, a_k\cdot P_k \left( \epsilon_k(x_k,a_k) \prod_{1\leq i \leq n}^{i
      \neq k} \phi_i(x_i,a_i) \right) \Bigg|_{\substack{x_i = x \\ a_i = 0}}
\end{equation*}
must vanish up to total derivatives when the Fierz equations for $\phi_i$ and $\epsilon_k$ are imposed. We
deduce that the corresponding vertex generating operator $\mathcal{V} \in \mathcal{R}$ satisfies
$[\mathcal{V}, a_k \cdot P_k] \approx 0$ for $k = 1, \ldots , n$. The operators $a_k\cdot P_k$ commute with
$s_{ij}$, hence, $[\Delta\mathcal{V} , a_k \cdot P_k] \approx 0$ for any product $\Delta$ of Mandelstam
variables, and since the ideal $\mathcal{I}_S + \mathcal{I}_D$ is gauge invariant, we find that
\begin{equation*}
  \left[Q_\mathcal{V} (y_{ii+1} , Y_i , s_{ij}) , a_k \cdot P_k \right] \approx 0\,.
\end{equation*}
Using
\begin{equation*}
  [y_{ii+1}, a_k \cdot P_k] = \delta_{ik} s_{i i +1}\,,  \qquad\qquad [Y_i , a_k \cdot P_k] = 0\,,
\end{equation*}
this reduces to
\begin{equation}
  s_{kk+1}\partial_{y_{kk+1}} Q_\mathcal{V} (y_{ii+1} , Y_i , s_{ij}) \approx 0\,,\label{GIC}
\end{equation}
where $Y_i$ is now treated as an independent variable: $\partial_{y_{k\,k+1}}Y_i=0$.

Note that all remaining equivalence relations in $\mathcal{I}_R$ are gauge invariant, hence, the generators of
$\mathcal{I}_R$ can be chosen to be polynomials only in $Y_i$ and $s_{ij}$. We conclude that because of
Eq.~(\ref{GIC}), $Q_\mathcal{V}$ can be chosen to be independent of $y_{ii+1}$.

\section{Parity-Odd Vertices}

So far, we only discussed parity-even deformations. The most general form of a \textit{parity-odd $n$-point
  vertex $\mathcal{L}_n$} is also given by Eq.~(\ref{eq:L^n}), but with the vertex generating operator
$\mathcal{V}$ replaced by a linear combination $\mathcal{W}$ of operators $\mathcal{V} \cdot B_{IJK}$, where
$\mathcal{V} \in \mathcal{R}$ and
\begin{equation*}
  B_{IJK} = \epsilon_{\mu\nu\rho} b^\mu_I b^\nu_J b^\rho_K,\qquad I,J,K = 1,\ldots,2n
\end{equation*}
contains a single epsilon tensor. Let $s_3$ be the $3\times3$ matrix that consists of the first three rows and
columns of $\mathcal{S}$. Then, $\det s_3 = 2s_{12}s_{13}s_{23}$ is a product of Mandelstam variables and
\begin{equation*}
  \det s_3 \cdot B_{IJK} = \frac{1}{6} 
  \left(
    \mathcal{B}_{1I} \mathcal{B}_{2J} \mathcal{B}_{3K} \pm 5 \text{ terms}
  \right) \cdot B_{123}\,.
\end{equation*}
This relation is proved using $\det s_3 = (B_{123})^2$.

We can now conclude along the lines of the previous sections: For a given parity-odd $n$-point vertex
$\mathcal{L}_n$, there exists a product $\Delta$ of Mandelstam variables, such that the corresponding vertex
generating operator $\mathcal{W}$ satisfies
\begin{equation*}
  \Delta \mathcal{W} \approx Q_\mathcal{W} (Y_i, s_{ij}) \cdot B_{123}\,,
\end{equation*}
where the polynomial $Q_\mathcal{W}$ is linear in the $Y_i$'s. The only additional input along this proof is
that $B_{123}$ is gauge invariant ($[B_{123},a_k \cdot P_k] = 0$).

\section{Final Steps}

Let us summarise: A Lorentz and gauge invariant parity-even $n$-point vertex $\mathcal{L}_n$ is given by
Eq.~(\ref{eq:L^n}) and there exists a product $\Delta$ of Mandelstam variables, such that the vertex generating
operator $\Delta\mathcal{V}$ is equivalent to a polynomial $Q_\mathcal{V}(Y_i, s_{ij})$, which is linear in each
$Y_i$. This means that there is no product $A_i^{\mu} A_i^{\nu}$ left, when we write $Q_\mathcal{V}$ in terms of
the operators $P_i^\mu$ and $A_i^\mu$. The equivalence relations do not change the number of those operators, so
this must also be true for $\Delta \mathcal{V}$. Finally, $s_{ij}$ (and thus, $\Delta$) only consist of the
operators $P_i^\mu$. We conclude that $\mathcal{V}$ cannot contain any product $A_i^{\mu} A_i^{\nu}$, meaning
that the corresponding $n$-point vertex $\mathcal{L}_n$, constructed via Eq.~(\ref{eq:L^n}), may only involve
fields whose spin is at most one. Note that for this argument, Eq.~(\ref{eq:main}) is essential.

For a parity-odd vertex, we use an analogous reasoning, except that $\Delta \mathcal{W}$ is equivalent to a
polynomial $Q_\mathcal{W}(Y_i, s_{ij})$ multiplied with $Q_{123}$. But since $Q_{123}$ does not contain any
$A_i^\mu$ operator, this does not alter the conclusion.

Finally, we find the extra equivalence relations $Y_i \approx 0$ for Chern-Simons fields $\phi_i$, which stem
from the corresponding free e.o.m. Hence, $\mathcal{L}_n$ may only contain massless scalars and Maxwell fields,
i.e. fields with propagating degrees of freedom. This completes the proof of the statement in the Introduction:
\textit{there are no independent vertices of order $n \geq 4$ that contain massless HS fields}.

\section{Conclusions}

We have shown in this Letter that in three dimensions gauge invariance strongly constrains the higher-order
interactions that involve massless fields. In particular, vertices that are independent of the cubic ones
can only contain scalars \footnote{Note that the scalar (Klein-Gordon) fields are also taken to be massless for
  simplicity, while the proof would go in the same way for scalar fields with arbitrary mass.} and Maxwell
fields, but no massless HS fields. Our argument should even apply when we include massive higher-spin fields in the set of propagating fields: Gauge invariance is so strong that it forbids massless HS fields to enter any independent higher-order vertex irrespective of the remaining field content of the theory. Furthermore, although the results were derived in flat space-time, they also hold for (A)dS (or
even any Einstein background) due to an argument given for the cubic vertices in \cite{Mkrtchyan:2017ixk}.

First of all, this implies that in any non-linear theory with HS spectrum, all higher-order vertices that only
include massless HS fields arise by the completion of the cubic ones to the full non-linear Lagrangian (as in
Yang-Mills theory or General Relativity). This has an interesting consequence in holography.
In~\cite{Mkrtchyan:2017ixk,Kessel:2018ugi} it was observed that cubic vertices satisfy triangle inequalities for
the spins. Our result implies that the only higher-order vertices are the completions of the cubic ones, and
they can be shown to satisfy polygon inequalities. This is in agreement with the CFT prediction
\cite{Fredenhagen:2018guf} and establishes a one-to-one map between bulk vertices and boundary correlators in
the context of $AdS_3/CFT_2$.

Secondly, the HS fields are known to admit a Chern-Simons description in three dimensions, in a first-order
formulation. The findings of this paper imply that all the fields that do not carry bulk propagating d.o.f.,
cannot participate in independent higher-order interactions. This is consistent with the statement on the
absence of higher-order self-interactions of Chern-Simons fields \cite{Barnich:1993vg} and may indicate an exact
equivalence of the Chern-Simons and metric descriptions of HS fields in the gauge sector. It is therefore
tempting to speculate that any non-linear action of massless HS fields without matter can be written in a
Chern-Simons form. Such an equivalence cannot extend to the matter sector though.

It would be interesting to extend the work to higher dimensions. In that case there will be independent
higher-order interactions for HS gauge fields. The classification of the vertices satisfying Eq.~(\ref{NE}) in
arbitrary dimensions will be given in \cite{SOK}.

\bigskip

\begin{acknowledgments}
  The authors are grateful to Andrea Campoleoni, Dario Francia, Maxim Grigoriev, Euihun Joung, Thorsten Schimannek and Arkady Tseytlin for useful discussions. The
  hospitality of the Erwin Schr{\"o}dinger International Institute for Mathematics and Physics during the
  program on ``Higher Spins and Holography'' is greatly appreciated.
\end{acknowledgments}

\end{document}